\documentclass{emulateapj}

\usepackage{bm}

\newcommand{\gcc}{\mbox{~g cm$^{-3}$}}
\newcommand{\beq}{\begin{equation}}
\newcommand{\eeq}{\end{equation}}
\newcommand{\bea}{\begin{eqnarray}}
\newcommand{\eea}{\end{eqnarray}}
\newcommand{\req}[1]{Eq.\ (\ref{#1})}
\newcommand{\mel}{m_\mathrm{e}}
\newcommand{\mpr}{m_\mathrm{p}}
\newcommand{\mH}{m_\mathrm{H}}

\newcommand{\omc}{\omega_\mathrm{ce}}
\newcommand{\omp}{\omega_\mathrm{cp}}

\shorttitle{Hydrogen atmospheres of magnetars}

\slugcomment{Received 2003 June 26; accepted 2003 September 9}

\begin{document}

\title{Equation of state and opacities for 
       hydrogen atmospheres \\
      of magnetars}

%%%  \author{A. Y. Potekhin$^1$, G. Chabrier$^2$}
%%% \affil{$^1$Ioffe Physico-Technical Institute,
%%%     Politekhnicheskaya 26, 194021 St.~Petersburg, Russia\\
%%%     $^2$Ecole Normale Sup\'erieure de Lyon
%%%     (CRAL, UMR CNRS No.\ 5574),
%%%     46 all\'ee d'Italie,\\ 69364 Lyon Cedex 07, France}

\author{Alexander Y. Potekhin\altaffilmark{1}}
\affil{Ioffe Physico-Technical Institute,
    Politekhnicheskaya 26, 194021 St.~Petersburg, Russia
\\{palex@astro.ioffe.ru}}

\and

\author{Gilles Chabrier}
\affil{Ecole Normale Sup\'erieure de Lyon
    (CRAL, UMR CNRS No.\ 5574), \\ % } \affil{
46 all\'ee d'Italie. 69364 Lyon Cedex 07, France
\\{chabrier@ens-lyon.fr}}

    \altaffiltext{1}{Also at the
    Isaac Newton Institute of Chile, St.~Petersburg Branch, Russia}
% ***********************************************************
\begin{abstract}
The equation of state and radiative opacities of 
partially ionized, strongly magnetized hydrogen plasmas,
presented in a previous paper [ApJ 585, 955 (2003)]
for the magnetic field strengths 
$8\times10^{11}$~G $\lesssim B \lesssim3\times10^{13}$
G, are extended to the field strengths 
$3\times10^{13}$~G $\lesssim B \leq 10^{15}$ G,
relevant for magnetars.
The first- and second-order thermodynamic functions
and radiative opacities are calculated 
and tabulated
for $5\times10^5$ K $\leq T\leq 4\times10^7$ K
 in a wide range of densities. 
We show that 
bound-free transitions give an important contribution to the 
opacities in the considered range
of $B$ in the outer neutron-star atmosphere layers. 
Unlike the case of weaker fields,
bound-bound transitions are unimportant.
\end{abstract}
\slugcomment{The Astrophysical Journal}
\maketitle

\section{Introduction} 
\label{sect-intro}
Neutron star atmospheres 
differ from the atmospheres of ordinary stars because of the high
gravity and magnetic fields. The majority of the known pulsars possess
magnetic fields  $B\sim10^{11}$--$10^{13.5}$ G \citep*{PSRcat}.  Most
popular models of  the soft gamma repeaters and anomalous X-ray pulsars
assume they are \textit{magnetars}, neutron stars with superstrong
magnetic fields $B\sim10^{14}$--$10^{15}$ G (e.g., \citealt{Thompson03}).

The properties of matter are strongly modified 
if the electron cyclotron energy $\hbar\omc=\hbar eB/\mel c$
exceeds 1 a.u.\ -- i.e., the field strength $B$
is higher than $B_0=\mel^2 c\, e^3/\hbar^3
= 2.3505\times10^9$~G
% where $\mel$ is the electron mass, $e$ the elementary charge,
% and $c$ the speed of light 
(for a review, see \citealt{Lai01}).
\citet*{Shib92} presented the first detailed
model of hydrogen atmospheres for neutron stars
with strong magnetic fields.
Later it was developed and used by many authors 
(e.g., \citealp*{Zavlin95,Pavlov95,PSZ96,Zane01,HoLai03,Ozel03};
and references therein). These works have played a valuable role in
assessing the observed spectra of the neutron
stars. 
Despite the recognition of the possibility
that the absorption by
 atoms can significantly contribute to the opacities,
the atoms were not included in the atmosphere model in a
thermodynamically consistent manner in the listed works.

Theoretical equation of state for partially ionized hydrogen plasma with
strong magnetic fields was developed by \citet*{PCS99}. The
nonperturbative effects of atomic thermal motion across the magnetic
field were taken into account and shown to be important. Based on this
theory, in a previous paper (\citealp{PC03}, hereafter Paper I)
 we performed extensive
calculations of the thermodynamic functions and calculated
monochromatic and Rosseland mean opacities, taking into account the
bound-bound and bound-free transitions. These calculations were done  at
$11.9\leq\log B [G] \leq 13.5$. In addition, we reevaluated the free-free
opacities, taking into account the motion of both interacting particles,
electron and proton, in arbitrary magnetic fields.

Here we report an extension of the previous calculations 
of the equation of state and opacities of a partially ionized hydrogen
plasma to higher field strengths, $13.5\leq\log B [G] \leq 15$. 

% **************************************************************
\section{Theoretical model}
\label{sect-input}

A major complication when incorporating bound species in the 
models of strongly magnetized neutron-star atmospheres
arises from the strong coupling between the center-of-mass motion
of the atom and the internal atomic structure.
If an atom rests without motion in a strong magnetic field,
there are two distinct classes of
its quantum states: at every value of the Landau quantum number $n$
and the magnetic quantum number $-s$ ($n\geq 0$, $s\geq -n$), 
there is one tightly bound
state, with
binding energy growing asymptotically (at $B\to\infty$)
as $(\ln B)^2$, 
and an infinite
series of loosely bound states (numbered by $\nu=0,1,\ldots$)
with binding energies approaching
the energies of a field-free H atom (e.g.,
\citealp{CanutoVentura}). 
The Landau number $n$ is not a ``good'' quantum number,
in the sense that
 the Coulomb interaction mixes different Landau orbitals
of a free electron, but this numbering is unambiguous and
convenient at $B\gg B_0$.
The atom is elongated:
its size along the magnetic field
$\bm{B}$ either decreases logarithmically (for 
the tightly bound states)
or remains nearly constant (for the loosely bound states), 
while the transverse radius decreases as $B^{-1/2}$.
The radiative transition rates 
are different for the three basic polarizations:
the linear polarization along the field and the two circular
polarizations in the transverse plane. 

This simplicity is destroyed when atomic motion is taken into account. 
The electric field, induced in the comoving  frame of reference, breaks
down the cylindrical symmetry. Separating the center-of-mass motion and
choosing an appropriate gauge of the vector potential 
\citep*{GD,VDB92,P94}, one comes to an effective one-particle
Schr\"odinger equation with an effective Hamiltonian depending on the
pseudomomentum $\bm{K}$, which is a conserved quantity related to the
center-of-mass motion in a magnetic field. If the component of $\bm{K}$
perpendicular to $\bm{B}$, $K_\perp$, is finite,
then not only $n$, but also $s$ is not a ``good'' quantum number;
nevertheless, it is convenient to retain the quantum level numbering by
${n}$, $s$, and $\nu$, specified above. In the adiabatic
approximation, widely used in the past \citep*{GD,IMS}, the
transverse part of the wave function is postulated to be the same as for
an electron without Coulomb interaction. We perform calculations without
this approximation, using the technique described by \citet{P94} for the
discrete  atomic spectrum and by \citet{PP97} for the continuum.
This technique uses the basis of Landau functions, which represent
the transverse parts of the wave functions of a free electron
 in a magnetic field and are the same
for the solutions of the Schr\"odinger and Dirac equations.
Therefore this technique can be used for the ordinary 
($\hbar\omc<\mel c^2$)
as well as
superstrong fields
($\hbar\omc>\mel c^2$), provided the atomic binding energy
$\ll\mel c^2$, which is always the case at $B\lesssim10^{15}$~G.

An atom moving across the strong magnetic field acquires a constant
dipole moment perpendicular to $\bm{B}$ and $\bm{K}$. Those radiative
transitions, which were dipole-forbidden  for an atom at rest because of
conservation  of the $z$-projection of the angular momentum, become
allowed and should be taken into account  in the atmosphere models. The
binding energies decrease with increasing  $K_\perp$. Asymptotically, at
large $K_\perp$, the binding energies tend to $e^3 B/(c K_\perp)$.

The equation of state for partially ionized hydrogen in  strong magnetic
fields was constructed and discussed by \citet{PCS99}. The treatment is
based on the free energy minimization. 
The free energy model is in essence a generalization of the 
nonmagnetic model of  \citet{SC91,SC92}
to the case of a strong magnetic field.
We consider a plasma composed of $N_\mathrm{p}$ protons, 
$N_\mathrm{e}$ electrons, $N_\mathrm{H}$ hydrogen atoms, and
$N_\mathrm{mol}$ molecules in a volume $V$, the number densities being
$n_j\equiv N_j/V$. The Helmholtz free energy is written as the sum
\beq
   F = F_\mathrm{id}^\mathrm{e} + F_\mathrm{id}^\mathrm{p} 
      + F_\mathrm{id}^\mathrm{neu}
       + F_\mathrm{ex}^\mathrm{C} + F_\mathrm{ex}^\mathrm{neu},
\label{Fren}
\eeq
where $F_\mathrm{id}^\mathrm{e}$, $F_\mathrm{id}^\mathrm{p}$, and
$F_\mathrm{id}^\mathrm{neu}$ are the free energies of ideal gases
of the electrons, protons, and neutral species, respectively, 
$F_\mathrm{ex}^\mathrm{C}$ takes into account the Coulomb plasma
nonideality, and $F_\mathrm{ex}^\mathrm{neu}$ is the nonideal
contribution which arises from interactions of bound species with
each other and with the electrons and protons. 
Ionization equilibrium is given by minimization of $F$ with
respect to particle numbers under the stoichiometric constraints,
keeping $V$ and
the total number $N_0$ of protons (free and bound)
constant. The latter number is determined by the total mass
density $\rho$:
$
  n_0 \equiv N_0/V \approx \rho/\mH,
$
where $\mH=\mel+\mpr$.
The formulae for each term in \req{Fren} are given in
\citet{PCS99} and in Paper I. 
The employed minimization technique is similar to
the technique presented by \citet{P96} for the nonmagnetic case.

Once the free energy is obtained, its derivatives 
over $\rho$ and $T$ and their combinations provide
the other thermodynamic functions.

The atomic number fraction $x_\mathrm{H}=n_\mathrm{H}/n_0$, 
evaluated in the
course of the free energy minimization, can be used in calculations of
atmospheric opacities. One should take into account that the strong
magnetic field affects the polarization properties of radiation
 (e.g., \citealp{Ginzburg}).
At photon energies $\hbar\omega$ much higher than
$
%   \hbar\omega_\mathrm{pl} =
        ({4\pi\hbar^2 e^2 n_\mathrm{e} 
                      / \mel} )^{1/2}
                      \approx 28.7\,\rho_0^{1/2}  \mathrm{~eV},
$
where 
% $\omega_\mathrm{pl}$ is the electron plasma frequency and 
$\rho_0\equiv\rho/(1\gcc)$,
radiation propagates in the form of two so-called \emph{normal modes}.
These modes have different polarization vectors
$\bm{e}_j$
and different absorption and scattering
coefficients, which depend on the angle $\theta_B$
between the propagation direction and $\bm{B}$
(e.g., \citealp*{Kam82}).
The two modes interact with each other via scattering.
\citet{GP73} formulated the radiative transfer
problem in terms of these modes.

Let the magnetic field be directed along the $z$-axis.
At a fixed photon frequency $\omega$,
the absorption opacity $\kappa_j^\mathrm{a}(\theta_B)$ 
in each mode $j$ and scattering opacities
$\kappa_{jj'}^\mathrm{s}(\theta_B)$ 
from mode $j$ into mode $j'$ can be
presented as (e.g., \citealp{Kam82})
\bea
&&\hspace*{-.7em}
   \kappa_j^\mathrm{a}(\theta_B) = \mH^{-1} \sum_{\alpha=-1}^1
     |e_{j,\alpha}(\theta_B)|^2 \,\sigma_\alpha^\mathrm{a},
\label{kappa-a}
\\&&\hspace*{-.7em}
 \kappa_{jj'}^\mathrm{s}(\theta_B) \!=\!\!
%     \left[ 
     {\frac34}
\!\!
  \sum_{\alpha=-1}^1 \!\!
     |e_{j,\alpha}(\theta_B)|^2 \,
  \nonumber\\&&\times
     {\sigma_\alpha^\mathrm{s}\over \mH}\int_0^\pi \!\!\!
       |e_{j',\alpha}(\theta_B')|^2\sin\theta_B'\,\mathrm{d}\theta_B',
%     \right],
\label{kappa-s}
\eea
where $\alpha=0,\pm1$,
 $e_{j,0}=e_{j,z}$ is the $z$-component of $\bm{e}_j$, and
$e_{j,\pm1}=(e_{j,x}\pm \mathrm{i} e_{j,y})/\sqrt{2}$
are the circular components.
The cross sections $\sigma_\alpha$ depend on 
$\omega$, but not on $j$ or $\theta_B$. 

The total scattering opacity from mode $j$ is
$\kappa_j^\mathrm{s}=\kappa_{j1}^\mathrm{s}+\kappa_{j2}^\mathrm{s}$,
and the total extinction opacity is
$\kappa_j=\kappa_j^\mathrm{a}+\kappa_j^\mathrm{s}$.
According to equations (\ref{kappa-a}), (\ref{kappa-s}),
we can write
\beq
   \kappa_j(\theta_B) = \sum_{\alpha=-1}^1 \!\!
     |e_{j,\alpha}(\theta_B)|^2 \,\hat\kappa_\alpha,
\eeq
where $\hat\kappa_\alpha$ ($\alpha=-1,0,1$) do not depend on
$\theta_B$. With a good accuracy, 
$\hat\kappa_\alpha \approx (\sigma_\alpha^\mathrm{a} + \sigma_\alpha^\mathrm{s})
/\mH$.

In a partially ionized atmosphere, the opacity is contributed
by electrons, ions, and bound species. 
The scattering cross section includes contributions from the
electrons and protons: 
$
   \sigma_\alpha^\mathrm{s} = \sigma_\alpha^\mathrm{s,e}
              + \sigma_\alpha^\mathrm{s,p}$.
With a good accuracy,
$\sigma_\alpha^\mathrm{s,e}$ and $\sigma_\alpha^\mathrm{s,p}$
are described by simple analytical formulae (e.g., \citealt{Pavlov95}).
The absorption cross section $\sigma_\alpha^\mathrm{a}$
includes contributions from absorption by plasma electrons and protons
(mainly by free-free transitions due to the electron-proton collisions,
 $\sigma_\alpha^\mathrm{ff}$),
transitions between discrete states of an atom
(bound-bound absorption, $\sigma_\alpha^\mathrm{bb}$) and
photoionization (bound-free, $\sigma_\alpha^\mathrm{bf}$).
Thus, for the hydrogen atmosphere, 
we can write
%% \bea&&
\beq
   \sigma_\alpha^\mathrm{a} \approx 
   x_\mathrm{H} (\sigma_\alpha^\mathrm{bb}+\sigma_\alpha^\mathrm{bf})
   + (1-x_\mathrm{H})\,\sigma_\alpha^\mathrm{ff} .
\label{kappa-mix}
\eeq
For the hydrogen atoms, electrons, and protons moving in a strong
magnetic field, the cross sections $\sigma_\alpha^\mathrm{bb}$
were studied by \citet{PP95},
$\sigma_\alpha^\mathrm{bf}$ by \citet{PP97},
and $\sigma_\alpha^\mathrm{ff}$ in Paper I.

\citet{HoLai03} presented convenient formulae for 
$|e_{j,\alpha}(\theta_B)|^2$
for a fully ionized, strongly magnetized electron-ion plasma 
(taking into account vacuum polarization). 
However, the bound species affect
the dielectric tensor of the medium and hence the 
polarization properties of the normal modes.
For a completely
neutral hydrogen gas in a strong magnetic field, 
polarization modes were studied 
by \citet{BulikPavlov},
but for a partially ionized gas the problem is not yet solved.
For the time being, we use
 the formulae of \citet{HoLai03}.
Since the neutral fraction
is typically small,
we expect that the resulting error should not be large for the 
effective Rosseland mean opacities presented in the tables.

% ***********************************************************
\section{Results}
% ***********************************************************
\subsection{Calculation of tables}
\label{sect-tables}
Since our model of the ionization equilibrium and equation of state 
is computationally expensive,
it is not possible to use our computer code ``on line'' in any practical
application. The alternative is to tabulate 
thermodynamic quantities covering the density, temperature, and
magnetic field domain of interest and 
to rely on an interpolation of the tabular values.
In Paper I, we presented the tables
which cover a range of $\rho$, $T$, and $B$
appropriate for most typical neutron stars, such as isolated radio
pulsars. We used 
the fitting formulae for the binding energies, quantum-mechanical sizes,
oscillator strengths,
 and electron collision widths derived by \citet{P98}.
Calculations at $B\gtrsim 10^{13.5}$ G were hampered by the absence
of the relevant fits.

\begin{figure}\epsscale{1}
\plotone{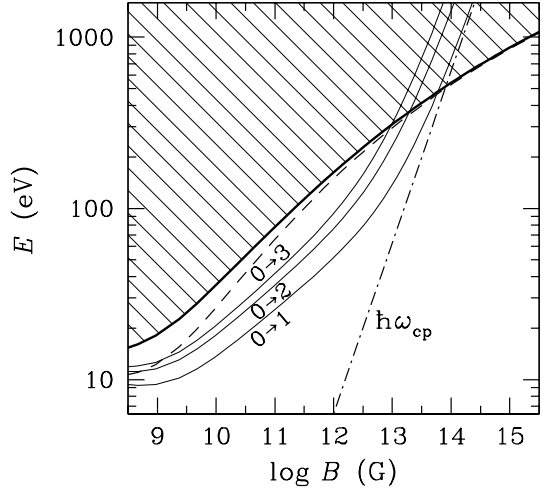}
\caption{Energies of radiative transitions 
for the H atom with $K_\perp=0$  
from the ground state to the lowest tightly bound states
 ($\nu=0$, $s=1,2,3$; light solid lines) and loosely bound state
 ($s=0$, $\nu=1$; dashed line); the photoionization threshold
 (heavy solid line); and the energy of the proton cyclotron
 resonance (dot-dashed line).
 The hatched region corresponds to the bound-free transitions.
\label{fig-en0tran}}
\end{figure}

In order to overcome this difficulty, in the present work
 we have calculated the
tables of the required atomic quantities as functions of the transverse
pseudomomentum $K_\perp$, using the computer code described by
\citet{P94}. A representative grid of $\approx 50$ equally spaced values
of $\log K_\perp$ has been used for every relevant set of atomic quantum
numbers and every considered field strength.
Handling of these tables is easier at superstrong than
at weaker fields, because the number of involved discrete states
is smaller. For example, there remain no bound states 
with $s=2$ 
and $s=1$ 
at $B>2\times10^{13}$ G and $B>6\times10^{13}$ G, respectively.
The latter point is illustrated by Fig.~\ref{fig-en0tran}, which shows
the energies of the most important atomic transitions at $K_\perp=0$
as functions of $B$. With increasing $B$,
the tightly bound levels $s=3,2,$ and 1 consecutively merge 
with the continuum and cease to contribute to the bound-bound spectrum;
instead, they appear as autoionization resonances. 
There still remain excited loosely bound states (with $s=0$ 
and any $\nu$) at any $B$,
but, at the typical neutron-star atmosphere densities,
most of them are destroyed by interactions with surrounding particles.

We use a cubic interpolation across these precalculated tables in our
computer code for obtaining the number fractions of atoms and molecules,
thermodynamic functions, and cross sections $\sigma_\alpha^\mathrm{bb}$.
As previously, we
employ the approximate expression
 for $\sigma_\alpha^\mathrm{ff}$ [Eq.~(51) in
Paper I] and numerical tables of $\sigma_\alpha^\mathrm{bf}$ as
functions of $K_\perp$ and $\omega$ (\S4.2 of Paper I).

In this way we have calculated and tabulated
thermodynamic functions and opacities
of the hydrogen plasma at $13.5\leq\log B\mbox{~[G]}\leq 15.0$
(preliminary results for $B=10^{14}$ G were presented
by \citealp*{CDP02}).
We have also updated the Fortran program for the cubic-polynomial 
interpolation of the tabulated data in the 3-parameter space
of $B$, $T$, and $\rho$.\footnote{%
The tables and the program are available at 
\url{http://www.ioffe.ru/astro/NSG/Hmagnet/}.}

As discussed by \citet{PCS99}, our model becomes less reliable 
at relatively low $T$ and high $\rho$, particularly because of formation 
of molecules and chains H$_n$, 
which are treated in an approximate manner.
In this domain, the partial number fractions and 
thermodynamic quantities are strongly model-dependent.
The stronger magnetic field, the higher 
the temperature below which the molecules and
chains dominate \citep{Lai01}.
Trying to avoid the domain of uncertainty, 
we shifted the lower bound on the
temperature in the tables, as compared to the case of weaker fields
(Paper I), from $2\times10^5$ K to $5\times10^5$ K.
On the other hand, the temperature at which the atoms become thermally
ionized increases with increasing $B$,
therefore
we have shifted the upper temperature bound in the tables
from $10^7$ K to $4\times10^7$ K.
It would be meaningless to consider still higher $T$,
because hydrogen undergoes efficient thermonuclear burning at such
temperatures \citep{Ergma}.
The third input parameter in our tables, in addition to $B$ and $T$,
is 
the ``astrophysical density parameter'' $R=\rho_0/T_6^3$,
where $T_6=T/10^6$ K.
In order to avoid the rather uncertain 
region of possible phase separation (see below),
we restricted the public tables by $\log R\leq3.4$,
although higher $R$ values were also considered.
This range of $R$ should be sufficient for modeling the atmospheres
of most neutron stars, except the unusually cold ones mentioned below.

\begin{figure*}\epsscale{1}
\plotone{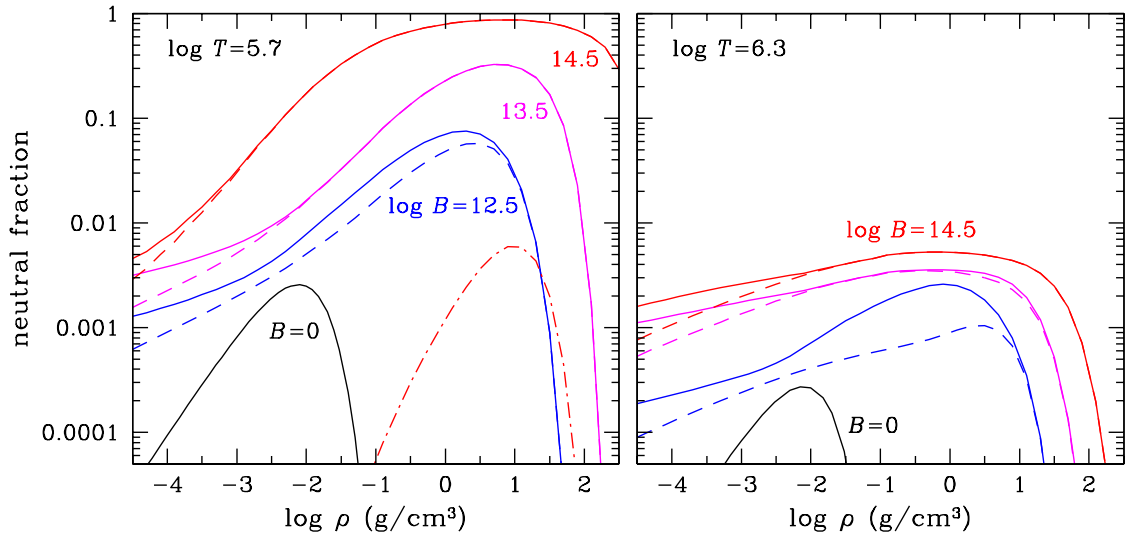}
\caption{Neutral fraction of the H atoms 
in all quantum states
(heavy lines)
and in the ground state (dashed lines) 
as function of density
at the magnetic field strength
$B=0$, $10^{12.5}$ G, $10^{13.5}$ G, and $10^{14.5}$ G.
The dot-dashed line shows the fraction of H$_2$ molecules
at the highest $B$. Left panel: $T=5\times 10^5$ K;
right panel: $T=2\times 10^6$ K.
\label{fig-ie-b}}
\end{figure*}

\begin{figure*}\epsscale{1.}
\plotone{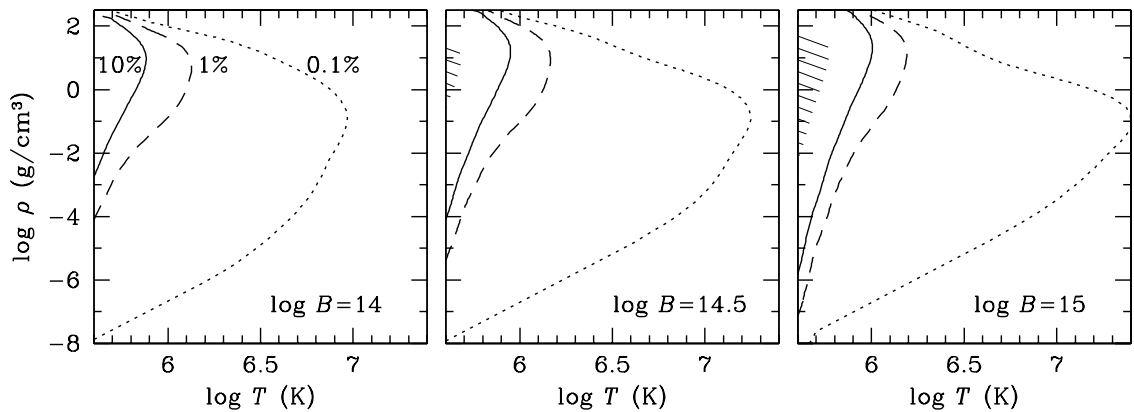}
\caption{Domains of partial ionization 
at $\log B\mbox{~[G]} =14.0$, 14.5, and 15.0. 
The contours delimit the domains where the atomic fraction
exceeds 0.1\% (dotted lines), 1\% (dashed lines), or 10\% (solid lines).
Hatched is the domain where the H$_2$ fraction exceeds 1\%.
\label{fig-phase3}}
\end{figure*}

Along with the pressure, our tables contain internal energy,
entropy, specific heat, and
the logarithmic derivatives of pressure  
with respect to $V$ and $T$.
Other second-order thermodynamic quantities
can be calculated using the Maxwell relations
\citep{LaLi-stat1}. 

The format of the tables is the same as in Paper~I.
In addition to the thermodynamic quantities, 
the tables contain the number fractions
of different chemical species, and also the effective Rosseland
mean opacities for diffusion of radiation along and across 
the magnetic field.

% ***********************************************************
\subsection{Ionization equilibrium and equation of state}
\label{sect-ioneq}
A strong magnetic field 
generally increases the fraction of bound species at a given $T$.
Figure \ref{fig-ie-b} illustrates this point.
In this figure, ionization equilibrium curves 
are plotted as functions of density for different values of $B$ and $T$.
At the lowest $T$ and highest $B$ shown in Fig.~\ref{fig-ie-b},
the molecular H$_2$ fraction (dot-dashed line on the left panel)
becomes significant at $\rho\gtrsim1\gcc$. At 
the same $T$ and $B$ and higher density
$\rho\sim100\gcc$,
the H$_2$ molecules disappear, but the H$_n$ chains and
clusters (not shown 
in the figure) become abundant. As mentioned above, our results
are not well justified under such conditions.

Figure \ref{fig-phase3} shows the domains of partial ionization
in the $T$--$\rho$ plane
at three values of $B$. With increasing $B$, the domains
where the atomic fraction is above a specified level
expand significantly towards higher temperatures.

\begin{figure}\epsscale{1}
\plotone{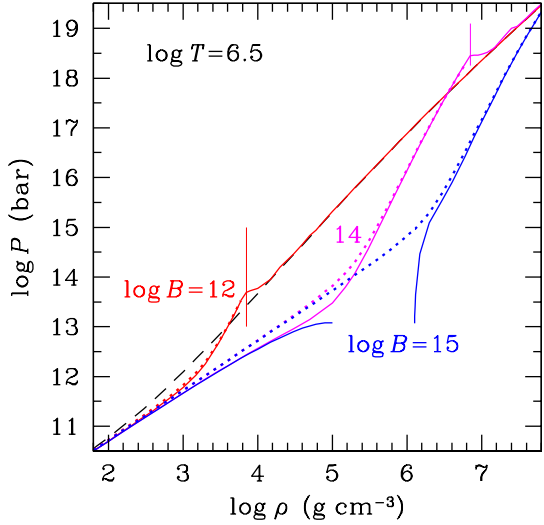}
\caption{Pressure isotherms ($T=10^{6.5}$ K)
of partially ionized hydrogen 
in the strong magnetic field ($B=10^{12}$~G, $10^{14}$~G,
and $10^{15}$~G; solid lines)
compared with the model of a fully ionized ideal
electron-proton plasma (dotted lines)
and with the case of a hydrogen plasma at $B=0$
(dashed line). 
The vertical lines correspond to the density above which
excited Landau levels become populated.
The instability region at $B=10^{15}$ G (the 
discontinuity of the lower solid curve)
corresponds to a hypothetical plasma
phase transition (see text).
\label{fig-eosmag}}
\end{figure}

Figure \ref{fig-eosmag} shows pressure $P$ 
as function of $\rho$ at $T=10^{6.5}$ K
and at different field strengths. At very low $\rho$,
we have nearly fully ionized, almost ideal nondegenerate
gas of electrons
and protons. The magnetic field does not affect
the equation of state in this case. With increasing density,
at $10^2\gcc\lesssim\rho\lesssim10^3\gcc$,
the electrons become degenerate at $B=0$, but remain nondegenerate
at $B\gtrsim10^{12}$ G. This explains why the dashed line in the figure
(pressure at $B=0$) is higher than the other lines in this regime.
Meanwhile, because of the partial recombination of atoms,
the solid curves ($P$ of the nonideal magnetized plasma)
are slightly lower than the dotted ones ($P$ of the ideal fully ionized
plasma).
At still higher density,
however, the atoms become pressure ionized: in this region
the increase of $P$ due to the increased number of free electrons and
protons competes with a negative nonideal contribution,
which is mainly due to the Coulomb term $F_\mathrm{ex}^\mathrm{C}$ 
in the free energy. 

At sufficiently high $B$ or low $T$, the
Coulomb interaction leads to a discontinuity, exemplified in
Fig.~\ref{fig-eosmag} by the line $B=10^{15}$ G. 
It occurs
at $T$ below the critical temperature
$T_\mathrm{c}$ and at $\rho$
around the critical density 
$\rho_\mathrm{c}\approx 143\,B_{12}^{1.18}\gcc$ \citep{PCS99},
where $B_{12}=B/10^{12}$~G.
This phase transition is directly related to the 
magnetic field assisted separation between the condensed hydrogen
and vapor, discussed by \citet{LS97}.
Note that the density of the condensed phase at zero pressure
and zero temperature,
estimated by \citet{LS97} as 
$\rho_\mathrm{s}\approx 560\,B_{12}^{6/5}\gcc$,
scales with $B$ similarly to $\rho_\mathrm{c}$. 
At $B=10^{15}$ G, this estimate gives 
$\rho_\mathrm{s}\approx2.2\times10^6\gcc$, whereas 
we have numerically $\rho_\mathrm{s}\approx1.7\times10^6\gcc$
in Fig.~\ref{fig-eosmag}.
However, in general, the validity of the free-energy models
in the framework of the chemical picture of plasmas 
is questionable near the plasma phase transition.
In order to study the matter properties near the phase transition, 
one needs to consider accurately
the interaction of H$_n$ molecules in the gas phase
and their interaction with the surface of the condensed phase. 
This problem goes beyond the scope of the present paper,
although it certainly deserves a future study.
If this phase separation is real, it can be important for
interpretation
of the emission of neutron stars with unusually low $T$ and/or high $B$,
which in this case do not possess an optically thick atmosphere
above the condensed surface. 
This hypothesis was suggested by \citet{LS97};
a possible example of such ``naked neutron star'' has been
recently considered by \citet*{Turolla03}.

\begin{figure*}\epsscale{1.}
\plotone{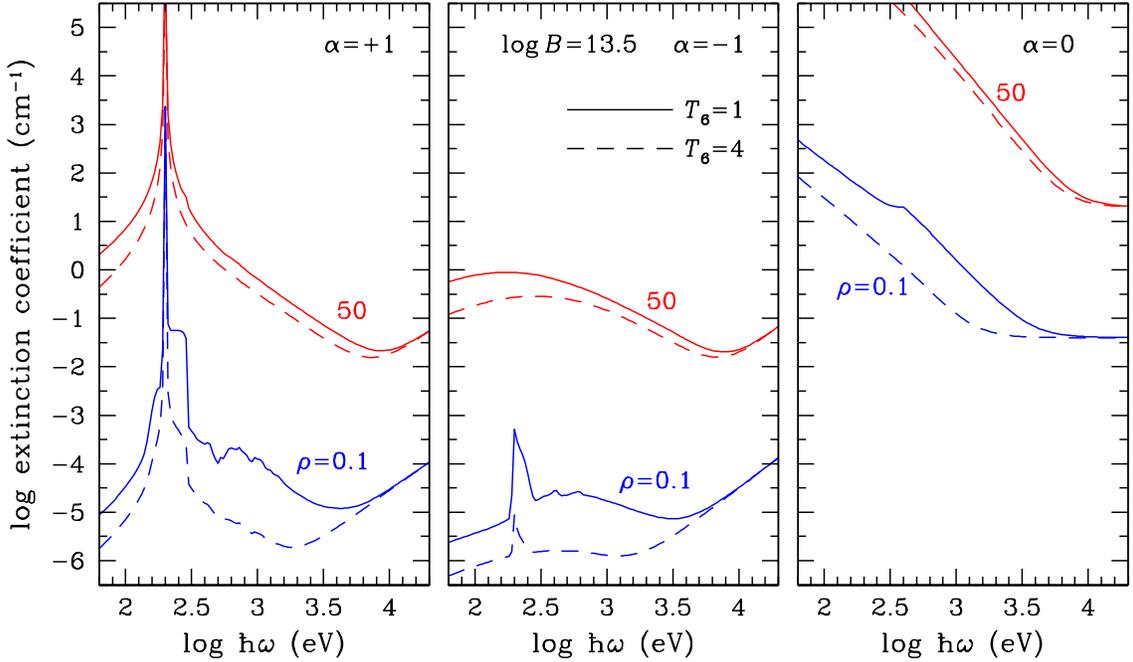}
\caption{Monochromatic
extinction coefficients for the three basic polarizations,
$k_\alpha=\rho\hat\kappa_\alpha$,
$\alpha=0,\pm1$,
at $\rho=0.1\gcc$ or $50\gcc$, $T=10^6$ K (solid lines)
or $4\times10^6$ K (dashed lines), $B=10^{13.5}$ G.
\label{fig-f13_5}}
\end{figure*}

\begin{figure*}\epsscale{1.}
\plotone{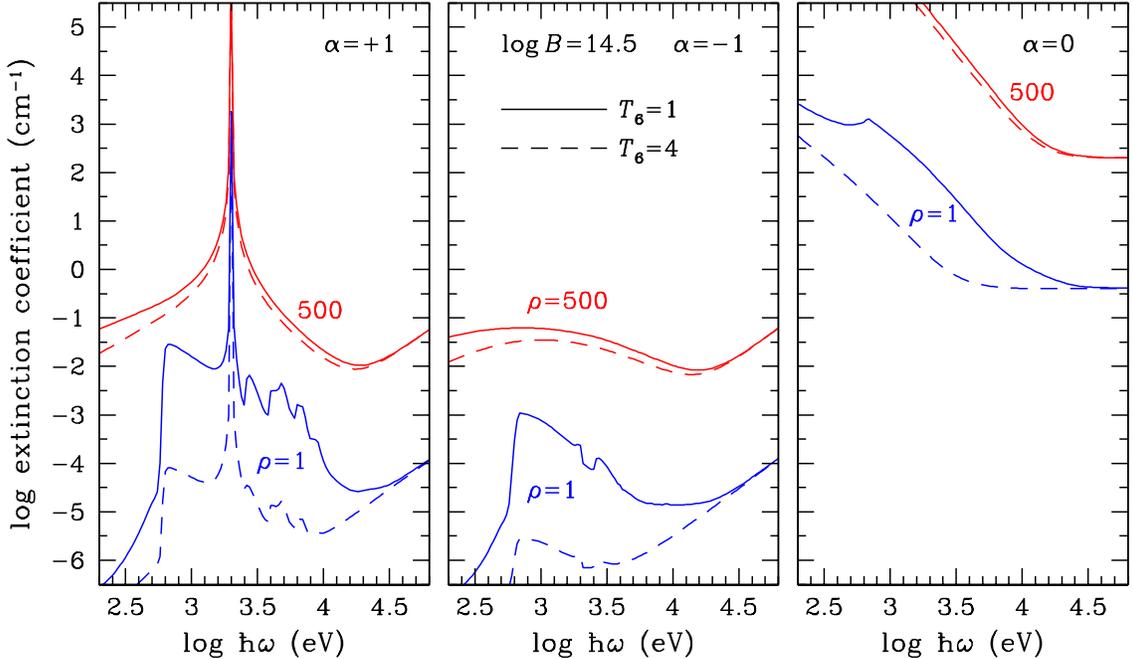}
\caption{The same as in Fig.~\ref{fig-f13_5}, but
for $\rho=1\gcc$ or $500\gcc$, $B=10^{14.5}$ G.
\label{fig-f14_5}}
\end{figure*}

% **********************************************************
\subsection{Opacities}
\label{sect-opac}
Figure \ref{fig-f13_5} shows the monochromatic extinction coefficients
$k_\alpha=\rho\hat\kappa_\alpha$ for the three basic polarizations
$\alpha=0,\pm1$ at $B=10^{13.5}$ G, $T_6=1$ or 4,
and $\rho=0.1\gcc$ or 50\gcc.
Figure \ref{fig-f14_5} demonstrates the same functions
for 10 times higher magnetic field and densities.
These densities correspond to the outer and inner layers of the
atmosphere, where the optical depth is of the order of 1 for ordinary
(high-opacity) and extraordinary (low-opacity)
modes of radiation, respectively 
(see, e.g., \citealt{HoLai03}).
At the higher densities, the opacities are featureless, except the
proton cyclotron resonance at 
$\hbar\omp=\hbar\omc \mel/\mpr=6.305\,B_{12}$ eV,
because the plasma at such density is pressure-ionized.
On the contrary, absorption features are clearly visible
at the lower density. 

At the lower field strength
(Fig.~\ref{fig-f13_5}), one sees the broad absorption feature for
$\alpha=\pm1$ (the left and central panels) due to the 
$s=0\to 1$
bound-bound transition at $\hbar\omega\lesssim0.3$ keV.
For $\alpha=+1$ (the left panel), this absorption merges with the strong
proton cyclotron free-free resonance at lower energies and with the
photoionization continuum at higher energies. This situation is rather
special, because at this field strength  $\hbar\omp$ is rather close to
the photoionization threshold, whereas the limiting energy for the
bound-bound transition falls in between (see Fig.~\ref{fig-en0tran}).
Both the photoionization edge and the absorption line are considerably
broadened because of the energy shifts due to
the atomic motion (the magnetic broadening; see \citealt{PP95} and
\citealp{PP97}),
therefore they effectively merge together.
On the other hand, the free-free absorption coefficient 
at the lower $T$ (the solid curve)
reveals the bumps 
at $\hbar \omega \lesssim1$ keV, which arise from
 the opening of  
ionization channels with higher $s$ values at higher
energies, and from autoionization resonances just below these partial
ionization thresholds \citep{PP97}.
On the right panel ($\alpha=0$, the polarization
vector along $\bm{B}$), one notices only the photoionization edge
at relatively small $\rho$ and $T$ (the lower solid line).

At the higher field strength (Fig.~\ref{fig-f14_5}), all transitions
changing $s$ belong to the continuum (see Fig.~\ref{fig-en0tran}) and
form a series of resonances on the photoionization profile at the
energies above $\hbar\omp$, particularly visible for $\alpha=+1$ at low
$\rho$ and $T$ (the lower solid line on the left panel). Near the bottom
of the photosphere ($\rho=500\gcc$), the bound states are destroyed and
do not contribute to the spectrum.

% ************************************************************
\section{Conclusions and outlook}
% ************************************************************
We have calculated the equation of state of 
fully and partially ionized hydrogen plasmas in a wide range
of densities, temperatures, and magnetic fields
expected for photospheres of the magnetars.
We have also calculated monochromatic radiative opacities for three
basic polarizations.
The first- and second-order thermodynamic functions, 
non-ionized fractions,
and effective Rosseland mean opacities
are published in the electronic form.

These results can be useful for
construction of the atmosphere models and calculation of 
the spectra of outgoing radiation
from warm neutron stars with superstrong magnetic fields.
This work is under way; some results have been published 
by \cite{Ho-COSPAR,hoetal03}. However, there remain
unsolved problems which hamper an immediate application of our 
present results to modeling accurate and reliable outgoing spectra of 
magnetars with hydrogen atmospheres. First, 
vacuum polarization alters the dielectric tensor of the medium
and polarization of photon modes (see \citealt{PG84}), 
which is important for magnetar atmospheres. 
In particular, the ``vacuum resonance'' 
phenomenon may lead, under certain conditions,
 to an adiabatic mode conversion
and ``mode collapse''
(\citealp{HoLai03}, and references therein).
In the latter case, the description in terms of the two modes
fails, and one has to solve more general equations for
the evolution of electromagnetic waves \citep{LaiHo03}.
Second, as noted in \S\ref{sect-input},
the presence of the bound species also changes 
the dielectric properties of the medium and the normal mode
polarization, which necessitates derivation of
a polarization tensor for partially ionized, strongly
magnetized plasmas.
Third, for relatively cold neutron stars with strong magnetic fields,
the phase separation mentioned in \S\ref{sect-ioneq} should be studied
in a thermodynamically consistent way. We are planning
to study these problems in near future.

      \acknowledgements
We thank Dong Lai for useful discussions, 
and Yura Shibanov and Wynn Ho for valuable remarks. 
A.P.\ gratefully acknowledges hospitality
of the theoretical astrophysics group 
at the Ecole Normale Sup\'erieure de Lyon
and the Astronomy Department of Cornell University.
The work of A.P.\ is supported in part by RFBR grants 
02-02-17668 and 03-07-90200.


\begin{thebibliography}{} 

\bibitem[Bulik \& Pavlov(1996)]{BulikPavlov}
Bulik, T., \& Pavlov, G. G. 1996,
ApJ 469, 373

\bibitem[Canuto \& Ventura(1977)]{CanutoVentura}
Canuto, V., \& Ventura, J. 1977,
%% ``Quantizing Magnetic Fields in Astrophysics,''
Fundam.\ Cosm.\ Phys., 2, 203

\bibitem[Chabrier et al.(2002)Chabrier, Douchin, \& Potekhin]{CDP02}
Chabrier, G., Douchin, F., \& Potekhin, A. Y. 2002,
J.\ Phys.: Condensed Matter, 14, 9133

\bibitem[Ergma(1986)]{Ergma}
Ergma, E. 1986, Sov.\ Sci.\ Rev.\ E:
Astrophys.\ Space Phys., 5, 181

\bibitem[Ginzburg(1970)]{Ginzburg}
Ginzburg, V. L. 1970,
The Propagation of Electromagnetic Waves in Plasmas,
2nd ed. 
%AA% (Pergamon, London)
        (London: Pergamon)

\bibitem[Gnedin \& Pavlov(1973)]{GP73}
Gnedin, Yu. N., \& Pavlov, G. G. 1973,
Zh.\ Eksper.\ Teor.\ Fiz., 65, 1806
[1974, Sov.\ Phys.--JETP, 38, 903]

\bibitem[Gor'kov \& Dzyaloshinskii(1967)]{GD}
Gor'kov, L. P., \& Dzyaloshinskii, I. E., 1967,
Zh.\ Eksper.\ Teor.\ Fiz., 53, 717 
[1968, Sov.\ Phys.--JETP, 26, 449]

\bibitem[Ho \& Lai(2003)]{HoLai03}
Ho, W. C. G., \& Lai, D., 2003,
MNRAS, 338, 233

\bibitem[Ho et al.(2003a)]{Ho-COSPAR}
Ho, W. C. G., Lai, D., Potekhin A. Y., \& Chabrier G. 2003a,
Adv.\ Sp. Res., in press (astro-ph/0212077)

\bibitem[Ho et al.(2003b)]{hoetal03}
Ho, W. C. G., Lai, D., Potekhin A. Y., \& Chabrier G. 2003b,
ApJ, accepted (astro-ph/0309261)

\bibitem[Ipatova et al.(1984)Ipatova, Maslov, \& Subashiev]{IMS}
Ipatova, I. P., Maslov, A. Y., \& Subashiev, A. V. 1984,
Zh.\ Eksper.\ Teor.\ Fiz., 87, 1804
[Sov.\ Phys.--JETP, 60, 1037]

\bibitem[Kaminker et al.(1982)Kaminker, Pavlov, \& Shibanov]{Kam82}
Kaminker, A. D., Pavlov, G. G., \& Shibanov, Yu. A. 1982,
Ap\&SS 86, 249

\bibitem[Lai(2001)]{Lai01}
Lai, D. 2001,
Rev.\ Mod.\ Phys., 73, 629

\bibitem[Lai \& Ho(2003)]{LaiHo03}
Lai, D., \& Ho, W. C. G. 2003,
ApJ 588, 962

\bibitem[Lai \& Salpeter(1997)]{LS97}
Lai, D., \& Salpeter, E. E. 1997,
ApJ, 491, 270

\bibitem[Landau \& Lifshitz(1993)]{LaLi-stat1}
Landau, L. D., \& Lifshitz, E. M. 1993,
Statistical Physics, Part 1
%AA% (Pergamon, Oxford)
       (Oxford: Pergamon)

\bibitem[Manchester et al.(2003)]{PSRcat}
Manchester, R. N., Hobbs, G. B., Teoh, A., \& Hobbs, M. 2003,
% ``A new pulsar catalogue,''
AJ, in preparation

\bibitem[{\"O}zel(2003)]{Ozel03}
{\"O}zel, F. 2003,
ApJ, 583, 402

\bibitem[Page et al.(1996)Page, Shibanov, \& Zavlin]{PSZ96}
Page, D., Shibanov, Yu. A., \& Zavlin, V. E. 1996,
% Temperature, distance and cooling of the Vela pulsar,
in MPE Report 263,
R\"ontgenstrahlung from the Universe,
ed.\ H. U. Zimmermann, J. E. Tr\"umper, \& H. Yorke
%AA% (MPE, Garching) 173 
       (Garching: MPE), 173 

\bibitem[Pavlov \& Gnedin(1984)]{PG84}
Pavlov, G. G., \& Gnedin, Yu. N. 1984,
Sov.\ Sci.\ Rev.\ E:
Astrophys.\ Space Phys., 3, 197

\bibitem[Pavlov \& Potekhin(1995)]{PP95}
Pavlov, G. G., \& Potekhin, A. Y. 1995,
ApJ, 450, 883

\bibitem[Pavlov et al.(1995)]{Pavlov95}
Pavlov, G. G., Shibanov, Yu. A., Zavlin, V. E., \& Meyer, R. D. 1995,
% ``Neutron Star Atmospheres'',
in NATO ASI Ser. C 450,
The Lives of the Neutron Stars,
ed. M. A. Alpar, \"U. Kizilo\u{g}lu, \& J. van Paradijs
%AA% (Kluwer, Dordrecht), 71
       (Dordrecht: Kluwer), 71

\bibitem[Potekhin(1994)]{P94}
Potekhin, A. Y. 1994,
J.\ Phys. B, 27, 1073

\bibitem[Potekhin(1996)]{P96}
Potekhin, A. Y. 1996,
Phys.\ Plasmas, 3, 4156

\bibitem[Potekhin(1998)]{P98}
Potekhin, A. Y. 1998,
J.\ Phys. B, 31, 49

\bibitem[Potekhin \& Chabrier(2003)]{PC03}
Potekhin, A. Y., \& Chabrier, G. 2003,
ApJ, 585, 955 (Paper I)

\bibitem[Potekhin \& Pavlov(1997)]{PP97}
Potekhin, A. Y., \& Pavlov, G. G. 1997,
ApJ, 483, 414

\bibitem[Potekhin et al.(1999)Potekhin, Chabrier, \& Shibanov]{PCS99}
Potekhin, A. Y., Chabrier G., \& Shibanov, Yu. A. 1999,
Phys.\ Rev. E 60, 2193

\bibitem[Saumon \& Chabrier(1991)]{SC91}
Saumon, D., \& Chabrier, G. 1991,
Phys.\ Rev. A 44, 5122

\bibitem[Saumon \& Chabrier(1992)]{SC92}
Saumon, D., \& Chabrier, G. 1992,
Phys.\ Rev. A 46, 2084

\bibitem[Shibanov et al.(1992)]{Shib92}
Shibanov, Yu. A., Pavlov, G. G., Zavlin, V. E., \& Ventura, J.
1992, A\&A 266, 313

\bibitem[Thompson(2003)]{Thompson03}
Thompson, C. 2003,
Mem.\ Soc.\ Astron.\ It., 73, 477

\bibitem[Turolla et al.(2003)]{Turolla03}
Turolla, R., Zane, S., \& Drake, J. 2003,
ApJ, accepted (astro-ph/0308326)

\bibitem[Vincke et al.(1992)Vincke, Le Dourneuf, \& Baye]{VDB92}
Vincke, M., Le Dourneuf, M., \& Baye, D. 1992,
J.\ Phys. B, 25, 2787

\bibitem[Zane et al.(2001)]{Zane01}
Zane, S., Turolla, R., Stella, L., \& Trevis, A., 2001,
ApJ, 560, 384

\bibitem[Zavlin et al.(1995)]{Zavlin95}
Zavlin, V. E., Pavlov, G. G., Shibanov, Yu. A., \& Ventura, J. 1995,
A\&A 297, 441

\end{thebibliography}
\end{document}